# Experimental demonstration of negative index of refraction

Jiangfeng Zhou[1], Thomas Koschny[2,3], Lei Zhang[1], Gary Tuttle[1] and Costas M. Soukoulis[2,3]

[1]*Department of Electrical and Computer Engineering and Microelectronics Research Center, Iowa State University, Ames, Iowa 50011*
[2]*Ames Laboratory and Department of Physics and Astronomy, Iowa State University, Ames, Iowa 50011*
[3]*Institute of Electronic Structure and Laser – FORTH, and Dept. of Materials Science and Technology, University of Crete, Greece*

**Abstract**

We introduce an improved and simplified structure made of periodic arrays of pairs of H-shaped metallic wires that offer a potentially simpler approach in building negative-index materials. Using simulations and microwave experiments, we have investigated the negative-index *n* properties of these structures. We have measured experimentally both the transmittance and the reflectance properties and found unambiguously that a negative refractive index with Re(n)<0 and Im(n) <Re(n). The same is true for ε and μ. Our results show that H-shaped wire pairs can be used very effectively in producing materials with negative refractive indices. © *2006 American Institute of Physics*
[ **DOI**: 10.1063/1.2208264,  **PACS** numbers: 41.20.Jb, 42.25.Bs, 42.70.Qs, 78.20.Ci

The first demonstration of a Left-Handed (LH) material [1,2] by the UCSD group, in 2000, following the work by Pendry *et. al.* [3,4], started a field of structured materials to create electromagnetic response not available in naturally occurring materials. This LH materials [5-8] made use of an array of conducting, nonmagnetic ring-shaped local resonators elements to achieve a negative effective permeability, $\mu(\omega)$, and an array of conducting continuous wires to achieve a negative effective permittivity, $\varepsilon(\omega)$; the simultaneous combination of which had never before been observed in any previously known material. LH materials display unique "reversed" electromagnetic properties, as discussed by Veselago [9] long before such materials existed.

In the past few years there has been ample proof for the existence of Negative Index Materials (NIMs) in the GHz frequency range [1-8]. Most NIM implementations to date have utilized the topology proposed by Pendry, consisting of split-ring resonators (SRRs) (rings with gaps, providing the negative μ) and continuous wires (providing the negative ε). A lot of groups were able to fabricate [1,2,5-8] NIMs with an index of refraction *n* = -1 and losses of less than 1 dB/cm. Recently different groups observed indirectly [10-13] negative $\mu$ at the THz region. In most of the THz experiments [10-12] only one layer of SRRs was fabricated on a substrate and the transmission, *T*, was measured only for propagation perpendicular to the plane of the SRRs, exploiting the coupling of the electric field to the magnetic resonance of the SRR via asymmetry [11]. This way it is not possible to drive the magnetic permeability negative. One reason is that is very difficult to measure with the existing topology of SRRs and continuous wires both the transmission, *T*, and reflection, *R*, along the direction parallel to the plane of the SRRs.

So there is a need for alternative, improved and simplified designs that can be easily fabricated and experimentally characterized. Currently, there is much interest in pushing the frequency range for NIM behavior into the infrared and optical regions of the spectrum [14-16]. However, the currently observed negative index in the THz region [12] is actually due to significant imaginary parts of ε and µ which lead to a dominant imaginary part of *n* and result in a rapid attenuation of EM wave consequently, which make such kind of meta-material inapplicable.

In this letter, we report our investigations into wire-pair structures as alternatives to conventional SRR-based NIMs. The basic structure of a single unit cell of the NIM build from H-shaped wires is shown in Fig.1A. In the wire-pair arrangement, the conventional SRR is replaced with a pair of short parallel wires, which provide both negative magnetic and electric response; the continuous wires are not necessary. The short wire pair consist of a pair of metal patches separated by a dielectric spacer of thickness $t_s$. For an electromagnetic wave incident with a wave vector and field polarization as shown in Fig. 1A, the short wire pair will process not only a magnetic resonance resulting in a negative µ [15-17] but also an electric resonance with a negative ε simultaneously. The magnetic resonance originates from the anti-parallel current in the wire pair with opposite sign charge accumulating at the corresponding ends; the electric resonance is due to the excitation of parallel current in the wire pair with same sign charge accumulating at the corresponding ends of both wires. Repeating this basic structure periodically in the x-, y-, and z-directions would result in a NIM structure.

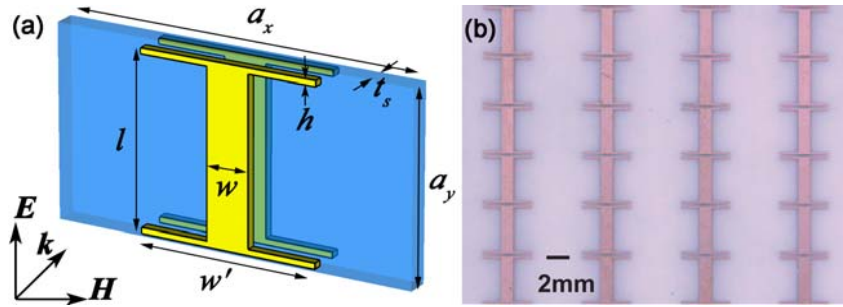

Figure 1. (a) Schematic representation of one unit cell of the wire-pair structure. (b) Photograph of fabricated microwave-scale wire-pair sample.

To examine the potential usefulness of the wire-pair structures as NIMs, we characterized the properties of the wire-pair of Fig. 1A using simulations and microwave measurements and then used these results to determine the expected properties of NIMs built from the wire-pair building blocks. Simulations of wire-pair structures were done with CST Microwave Studio (Computer Simulation Technology GmbH, Darmstadt, Germany), which uses a finite-difference time-domain method to determine reflection/transmission amplitudes of metallo-dielectric structures. In the simulations, the dielectric properties of the metal patches were described by a frequency-dependent Drude model. The detailed calculations were used to determine the reflection and the transmission coefficients from a single unit cell with periodic boundary condition in the direction perpendicular to the direction of the incident wave. Experimental transmission and reflection data were obtained by building and measuring microwave-frequency versions of the wire-pair structures. These were fabricated using Rogers 5880 printed circuit board stock with dielectric-layer thickness of 254 µm and listed relative dielectric constant of 2.53. The circuit board was coated on both sides with 10-µm thick layers of copper. The copper was formed in the wire-pair patterns using conventional photolithography techniques. For the samples reported here (both simulations and experiments), the width of the metal wires was 1 mm. The length and the width of the metal bars at each end of the wire pairs was 4mm and 0.2 mm respectively. The length of the short wire pairs was 4 mm, and the unit cell size was 4.2 mm x 8 mm x 2.274 mm. The total sample size was 9 x 17 x 1 unit cells, resulting in approximately square samples. A photograph of one side of a complete sample is shown in Fig. 1B. With these patterned dimensions on the printed-circuit board material, the resonances for NIM-behavior were expected to occur near 15.8 GHz

Transmission and reflection properties of a single-layer structure were measured over the frequency range of 14 – 18 GHz using a network analyzer (HP 8510) and a pair of standard gain horn antennas serving as source and receiver. In the transmission measurements, the microwaves were incident normal to the sample surface. This is a tremendous simplification relatively to the conventional SRRs and wires where the incident electromagnetic waves have to propagate parallel to the sample surface. With the conventional orientation of the SRRs, it is almost impossible to do these type of measurements at the THz region, since only single-layer samples are usually fabricated [10,11]. Transmission measurements were calibrated to the transmission between the horns with the sample removed. The reflection measurements were done by placing the source and receiving horns on the same side of the sample and bouncing the microwave signal off the sample. The source and receiver horns were each inclined with an angle of about 7.5° with respect to normal on the sample surface. The reflection measurement was calibrated using a sample-sized sheet of copper as reflecting mirror. In both measurements, the electric field of the incident wave was polarized parallel to the long dimension of the wires. (For perpendicular polarization, the transmission was nearly 100%, independent of frequency in the resonance region, and reflection was essentially zero.)

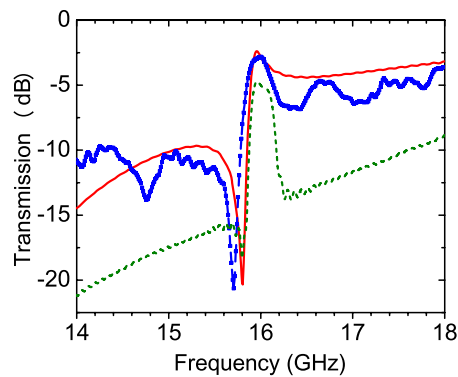

Figure 2. Simulated (red solid curve) and measured (blue dotted curve) transmission spectra for electromagnetic radiation incident on the wire-pair structures. The green dashed curve shows the simulated transmission spectrum for a 5-layers sample.

The calculated and measured transmission spectra are shown in Fig. 2. There is good qualitative agreement between simulations and measurements. To demonstrate the appearance of the expected left-handed transmission band more clearly than it is visible from the single unit cell spectra, we include the simulated transmission through five unit cells as the dashed line. The measured spectrum does show additional resonance peaks and valleys due to reflections between the receiving horn and the sample. Using the transmission and reflection results from a single layer, we can extract the effective refractive index that would result if a periodic multi-layer sample were built using the single-layer structure as a building block. The details of the numerical retrieval procedure have been described in detail elsewhere [18-20]. In performing the retrieval, we have assumed a z-direction size of the unit cell of 2.274 mm. Such that the individual wire-pair layers are separated from each other by a significant air space.

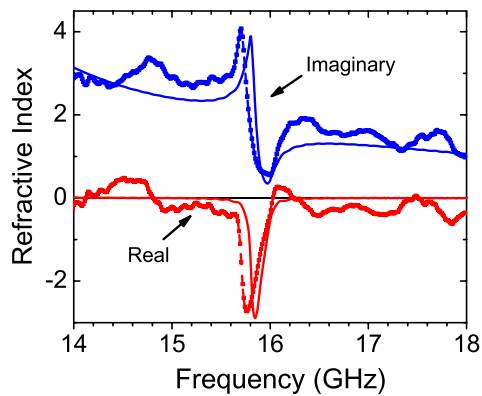

Figure 3. Extracted refractive index *n* of a periodic array of wire-pair unit cells, using the simulated (solid curves) and measured (dotted curves) transmission and reflection data. The red and blue curves show the real part of *n* and imaginary part of *n* respectively.

The extracted refractive index is shown in Fig. 3 and the extracted permittivity and permeability are shown in Fig. 4. The plots show that the real part of the permittivity is negative over most of the measured range. The real part of the permeability is negative over a resonance band above 15.8 GHz for both of the simulation (approximate 15.87-16.00 GHz) and the measurement (approximate 15.82-15.98 GHz). The extracted real part of the refractive index is negative [21-23] over a narrow band around 15.8 GHz for both of the simulations (15.59-16.17 GHz) and the experiments (15.67 -16.02 GHz), dipping as low as -2.66 using measured data and -2.86 from the simulation. The ratio of the imaginary part of n to the real part of n reaches 1/3 above the resonance, which means that we have left-handed propagation with ε, μ and n negative. The simplicity of the short-wire pair design and the alleviation of the need for continuous wires generating the negative effective permittivity is expected to facilitate scaling of the structure to achieve left-handed response well within the THz region. However the reader should be aware that straightforward geometric scaling of the present design is not possible as the behavior of the metals changes from lossy conductors in GHz to lossy, negative dielectrics in the higher THz region [24]. Our preliminary numerical results show that a negative index left-handed band with both, εand μ negative, and n'/n'' ratio better than 6 can be achieved combining the second order magnetic resonance with the first order electric resonance for Drude-model silver short-wire pairs at around 490 THz.

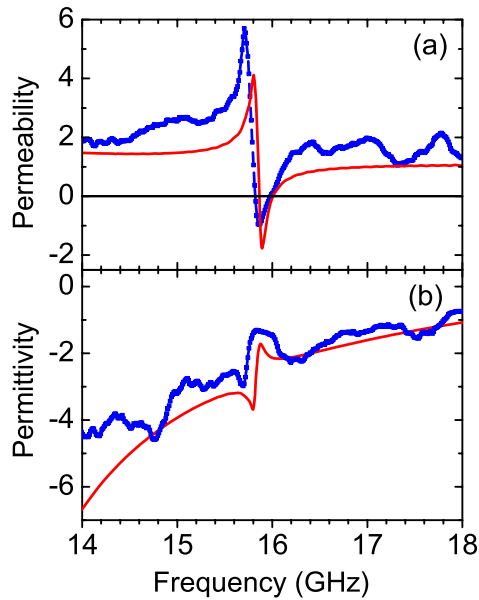

Figure 4 Extracted permittivity $\varepsilon$ (a) and permeability $\mu$ (b) of a periodic array of wire-pair unit cells, using the simulated (red solid curves) and measured (blue dotted curves) transmission and reflection data.

These results show clearly the viability of using short wire pairs to build negative-index materials. It is likely that modifications of the basic structure studied here may improve or alter the NIM properties. Also, wire-pair arrangements with significantly different geometries may lead to negative-index materials. The relative ease of fabricating wire-pair structures pairs may hasten the development of NIMs working at optical wavelengths.

We gratefully acknowledge the support of Ames Laboratory, which is operated by Iowa State University under contract No. W-7405-Eng-82, EU FET projects DALHM, PHOREMOST, METAMORPHOSE, and DARPA (Contract No. HR0011-05-C-0068).

**References**

1.  D. R. Smith, Willie J. Padilla, D. C. Vier, S. C. Nemat-Nasser, and S. Schultz, *Phys. Rev. Lett.* **84**, 4184 (2000).

2.  R.A. Shelby, D.R. Smith & S. Schultz, *Science* **292**, 77 (2001).

3.     J.B. Pendry, A.J. Holden, D.J. Robbins, W.J. Stewart, *IEEE Trans. MIT* **47**, 2075 (1999);  J.B. Pendry, A.J. Holden, W.J. Stewart, I. Youngs, *Phys. Rev. Lett*. **76**, 4773 (1996).

4.     J.B. Pendry, *Phys. Rev. Lett.* **85**, 3966 (2000).

5.     C.G. Parazzoli, R.B. Greegor, K. Li, B. E. C. Koltenbah, M. Tanielian, *Phys. Rev. Lett.* **90**, 107401 (2003); K. Li, S.J. McLean, R.B. Greegor, C.G. Parazzoli, M.H. Tanielian, *Appl. Phys. Lett*. **82**, 2535 (2003).

6.     M. Bayindir, K. Aydin, E. Ozbay, P. Markos, C.M. Soukoulis, *Appl. Phys. Lett*. **81**, 120 (2002); Aydin K, Guven K, Kafesaki M, Zhang L, Soukoulis CM, Ozbay E, *Optics Letters* **29**, 2623 (2004).

7.     E. Cubukcu, K. Aydin, E. Ozbay, S. Foteinopoulou, C.M. Soukoulis, *Nature* **423**, 604 (2003);  and E. Cubukcu, K. Aydin, E. Ozbay, S. Foteinopoulou, C.M. Soukoulis, *Phys. Rev. Lett*. **91**, 207401 (2003).

8.     P.V. Parimi, W.T.T. Lu, P. Vodo, S. Sridhar, *Nature* **426**, 404 (2003).

9.     V.G. Veselago, *Soviet Physics, USPEKHI* **10**, 509 (1968).

10.    T.J. Yen, W.J. Padilla, N. Fang, D.C. Vier, D.R. Smith, J.B. Pendry, D.N. Basov , X. Zhang, *Science* **303**, 1494 (2004).

11.    S. Linden, C. Enkrich, M. Wegener, J.F. Zhou, T. Koschny, C.M. Soukoulis, *Science* **306**, 1351 (2004).

12.     S. Zhang S, W.J. Fa, B.K. Minhas, A. Frauenglass, K.J. Malloy, S.R.J. Brueck , *Phys. Rev. Lett*. **94**, 37402 (2005); S. Zhang, W.J. Fan, N.C. Panoiu, K.J. Malloy, R.M. Osgood, S.R.J. Brueck. *Phys. Rev. Lett*. **95**,137404 (2005).

13.    N. Katsarakis, G. Konstantinidis, A. Kostopoulos, R.S. Penciu, T.F. Gundogdu, M. Kafesaki, E.N. Economou, T. Koschny, C.M. Soukoulis, *Optics Letters* **30**, 1348 (2005).

14.    V.A. Podlovsk, A.K. Sarychev, V.M. Shalaev, J. of Non. Opt. Phys. & Mat. **11**, 65 (2002)**;** V.A. Podlovsk, A.K. Sarychev, V.M. Shalaev, *Optics Express*, **11**, 735 (2003).

15.    G. Dolling, C. Enkrich, M. Wegener, J. F. Zhou, C. M. Soukoulis, and S. Linden, *Opt.Lett.*, **30**, 3198 (2005).

16.    V. M. Shalaev, W. Cai, U. Chettiar, H. K. Yuan, A.K. Sarychev, V.P. Drachev, A.V. Kildishev, *Opt.Lett*. **30**, 3356 (2005).

17.    J.F. Zhou, L. Zhang, G. Tuttle, Th. Koschny and C.M. Soukoulis, *Phys. Rev. B*, **73**, 041101(R) (2006).

18.    D. R. Smith, S. Schultz, P. Markos and C. M. Soukoulis, *Phys. Rev. B* **65**, 195104 (2002).

19.     D. R. Smith, D. C. Vier, Th. Koschny, and C. M. Soukoulis, *Phys. Rev. E* **71**, 036617 (2005).

20.     Th. Koschny, P. Markos, E. N. Economou, D. R. Smith, D. C. Vier, and C. M. Soukoulis, *Phys. Rev. B* **71**, 245105 (2005).

21. In lossy materials is possible to have the real part *n* to be negative, without having the real parts of ε and μ simultaneously negative. This is the case of the recent work of S. Zhang, W.J. Fan, N.C. Panoiu, K.J. Malloy, R.M. Osgood, S.R.J. Brueck. *Phys. Rev. Lett.* **95**,137404 (2005). This can happen if the imaginary parts of ε and μ are sufficiently large, because in a lossy material $n = n' + in''$, and we also have that $n = \varepsilon z$ and $z = \sqrt{\mu/\varepsilon}$. After some algebra we obtain that $n' = \varepsilon' z' - \varepsilon'' z''$ and $z = \sqrt{\frac{\mu'\varepsilon' + \mu''\varepsilon''}{\varepsilon^2} + i\frac{\mu''\varepsilon' - \mu'\varepsilon''}{\varepsilon^2}}$, so it's possible to have $n' < 0$, provided that $\varepsilon'' z'' > \varepsilon' z'$. In this scenario which occurs at the low-frequency side of the $n' < 0$ region in Fig. 3, however, the imaginary parts lead to dominant losses such that we have a transmission gap with some negative phase shift rather than LH transmission (with some losses). This type of negative *n* should not be considered LH behavior. In our experiments, although we have considerable imaginary parts, the behavior is still dominated by the negative real part of *n* at the high-frequency side where we find the LH behavior. As one can see from the Fig.3, we obtain *n'/n''=2* for experiment and *n'/n''=3* for simulation *at n'=-2*.

22. M.W. McCall, A. Lakhtakia & W.S. Weiglhofer, *European Journal of Physics*, **23**, 353 – 359 (2002).

23. R.A. Depine & A. Lakhtakia, *Microwave and Optical Technology Letters*, **41**, 315 – 316 (2004).

24. J.F. Zhou, T. Koschny, M. Kafesaki, E. N. Economou, J. B. Pendry, and C. M. Soukoulis, *Phys. Rev. Lett.*, **95**, 223902 (2005).